\documentclass[multphys,vecphys]{svmult}

\usepackage{makeidx}         
\usepackage{graphicx}        
\usepackage{multicol}        
\usepackage[bottom]{footmisc}

\makeindex             

\begin{document}

\title*{Globular Clusters and Galaxy Formation}
\author{Duncan A. Forbes}
\institute{Centre for Astrophysics \& Supercomputing, Swinburne
University, Hawthorn VIC 3122, Australia\\
\texttt{dforbes@swin.edu.au}
}
%
%
\maketitle


\begin{abstract}
We first discuss recent progress in using the Milky Way globular
cluster (GC) system as a `test-bed' for properties derived from 
integrated spectra and stellar population models. Standard
techniques may give rise to spuriously high alpha-element ratios
at low metallicities.
We then discuss evidence for early epoch (z $\ge$ 2) formation for
most GCs in galaxies today. Recent accretions of GCs (and their host
galaxy) make a small contribution but recent mergers form few if
any new GCs in today's elliptical galaxies. 
The early formation of metal-poor GCs and the 
bimodality seen in GC specific frequency requires a `truncation'
which may be due to reionization.

\end{abstract}

\section{Milky Way Globular Cluster System}
\label{sec:1}

The globular cluster (GC) system of our own Galaxy provides an
obvious test-bed for the integrated spectral 
properties that we can measure for
extragalactic GCs, such as age, metallicity and alpha-element
abundance. It also allows us to test the single stellar
population (SSPs) models that we rely on. Limitations on this
approach include the small number of Milky Way GCs
($\sim$150), strong extinction to some (particularly metal-rich
bulge GCs), a limited age range (most GCs are $>$ 10 Gyrs old)
and the difficulty of obtaining well calibrated, high S/N spectra.  
Some inroads on the latter have been made by Puzia et al. (2002),
Cohen et al. (1998) and Schiavon et al. (2005). From this combined
dataset we have formed a database of 48 distinct Milky Way GCs on
the Lick system with S/N $\sim$ 100. 

In Proctor, Forbes \& Beasley (2004) we showed that more reliable
results are obtained from using all available Lick lines than
H$\beta$ or H$\gamma$ alone. Recent fits to the combined dataset
by J. Mendel shows a good agreement with the CMD-derived ages and
metallicities of de Anglei et al. (2005). However, the trend for
alpha-elements with metallicity rises to unrealistically high
values at low metallicities (see Fig. 1). Given that high
resolution spectra of Carney and coworkers indicates a near
constant [E/Fe] = +0.3 for Milky Way GCs, we suspect this is an
SSP modelling effect. We have modified the base SSP model of Lee
\& Worthey (2005) to include the 
Houdashelt et al. (2002) formulation for
alpha-elements. The refit results are shown in Fig. 2 -- they
now reveal the expected trend. So although preliminary, this
approach shows promise. 

\begin{figure}
\centering
\includegraphics[height=10cm, angle=-90]{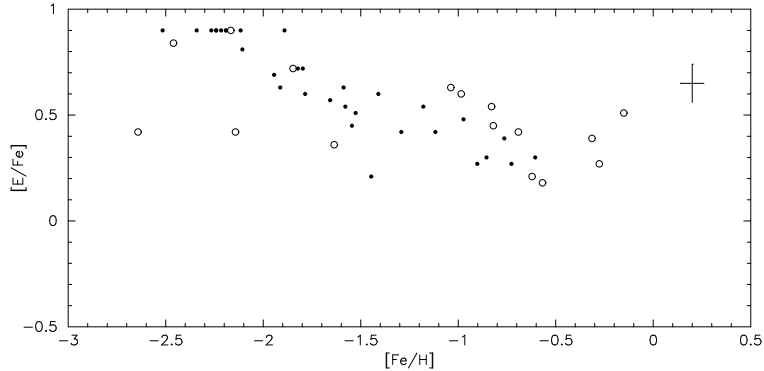}
\caption{Alpha-element ratio vs metallicity for Milky Way
GCs. Symbols represent different datasets. Fits were carried out
using Korn et al. (2005) SSP models. The rise to large alpha ratios at low
metallicities is not seen in high resolution spectra.
}
\label{fig:1}       
\end{figure}

\begin{figure}
\centering
\includegraphics[height=10cm, angle=-90]{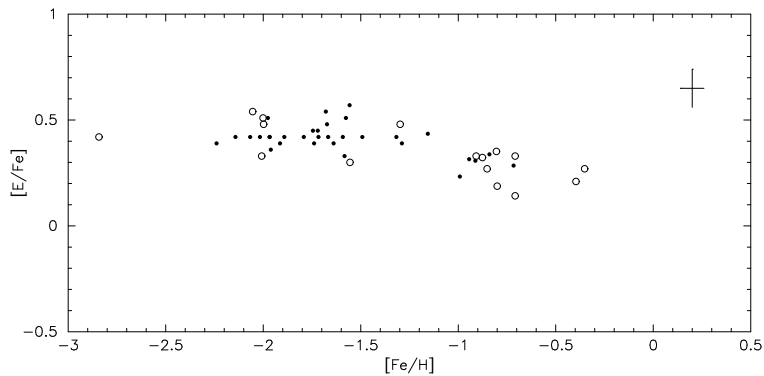}
\caption{Alpha-element ratio vs metallicity for Milky Way
GCs. Same as Fig. 1 except we used the base SSP model of Lee \&
Worthey (2005) modified by Houdashelt et al. (2002). 
Here the expected alpha-element
trend is found.   
}
\label{fig:1}       
\end{figure}

\section{Early Globular Cluster Formation}

Forbes, Brodie \& Grillmair (1997) and Beasley et al. (2002)
favoured a formation scenario in which GCs were formed in a
dissipative process in the early Universe. Both required a
mechanism to cutoff or truncate the formation of metal-poor GCs
(which formed first) before the second phase of metal-rich GC
formation occurred. Reionization now seems to be the `best bet'
process for this truncation (e.g. Santos 2003). If correct, a
fossil record of the epoch of reionization can be found by
examining the surface density profile of metal-poor GCs (Moore et
al. 2006; Bekki 2005). This prospect is particularly exciting
given the current constraints of z$_{reion}$ $\sim$ 6 from QSOs
and z$_{reion}$ $\sim$ 11 from the WMAP satellite.   

\begin{figure}
\centering
\includegraphics[height=10cm, angle=-90]{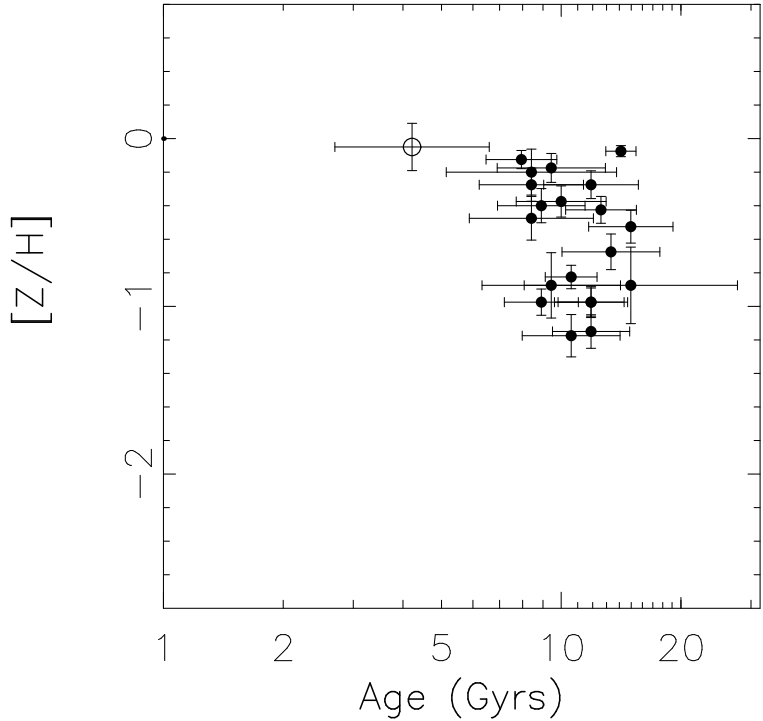}
\caption{Age-metallicity relation for NGC 4365 GCs. All sampled GCs
appear to be old. The one possible 
exception (open symbol) has a blue horizontal branch that causes
us to underestimate its age.
}
\label{fig:1}       
\end{figure}

The combination of truncation of metal-poor GC formation and a
galaxy mass-to-light ratio that varies with mass, can explain the
U-shaped distribution in GC specific frequency with host galaxy
luminosity (Forbes 2005; Bekki, Yahagi \& Forbes 2006). Such
bimodality in GC properties with galaxy mass may be explained by
the transition from hot to cold flows (Dekel \& Birnboim 2006).

The most direct way of probing the formation epoch of GCs is to
determine GC ages from high S/N spectra. The SAGES team have been
carrying out such work for several years with the Keck
telescope. It appears that the vast bulk of GCs in elliptical
galaxies are old, i.e. ages $>$ 10 Gyrs. An interesting, and
controversial case, is NGC 4365 in the Virgo cluster. In Fig. 3
we show the age-metallicity relation (AMR) for GCs in this
galaxy. Ages and metallicities have been fit using the multi-line
method of Proctor, Forbes \& Beasley (2004) by M. Pierce with 
indices from Brodie et al. (2005). The AMR shows that our
sample GCs are all old, i.e. forming at redshifts z $\ge$ 2. One
possible exception (open symbol) has a clear Calcium H+K
`inversion' in its spectrum indicative of a blue horizontal
branch; so in this case the age is underestimated and the
metallicity overestimated.

\section{Recent Accretions}

\begin{figure}
\centering
\includegraphics[height=10cm, angle=-90]{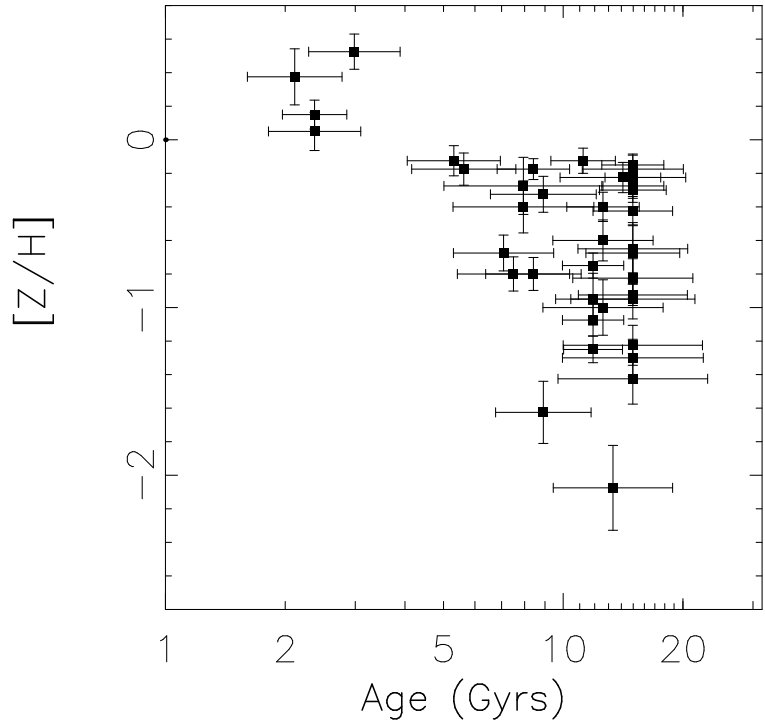}
\caption{Age-metallicity relation for NGC 4649 GCs. Most GCs are
old, however 4 GCs appear to be quite young (2--3 Gyrs). These
GCs are not all centrally located and we speculate that they may have
been accreted from a dwarf galaxy. 
}
\label{fig:1}       
\end{figure}

In Forbes et al. (2004) we presented AMRs for the GCs associated
with the accreted Sgr and Canis Major dwarf galaxies. We further
suggested that dwarf galaxy accretion contributed to the
increased age dispersion at intermediate metallicities of the
Milky Way GC system, but overall the contribution of accreted
GCs was small. A rare example of a dwarf galaxy (and presumably
its GC system) in the process of being accreted by a larger
spiral galaxy was presented in {\it Science} by Forbes et
al. (2003).

A possible example of accreted GCs in a nearby elliptical galaxy
is NGC 4649 (Pierce et al. 2006). The AMR shown in Fig. 4
reveals that the bulk of the GCs are old, however there are 4 GCs
which appear to be young ($\sim$2--3 Gyrs) ages. Interestingly, these
GCs are not all centrally located or on the side of the galaxy
towards its nearest neighbour NGC 4647, but lie in different
parts of the galaxy. We speculate that they may have been
accreted from a gaseous dwarf galaxy, of which there is no current
sign.

\section{Recent Mergers}

\begin{figure}
\centering
\includegraphics[height=10cm, angle=0]{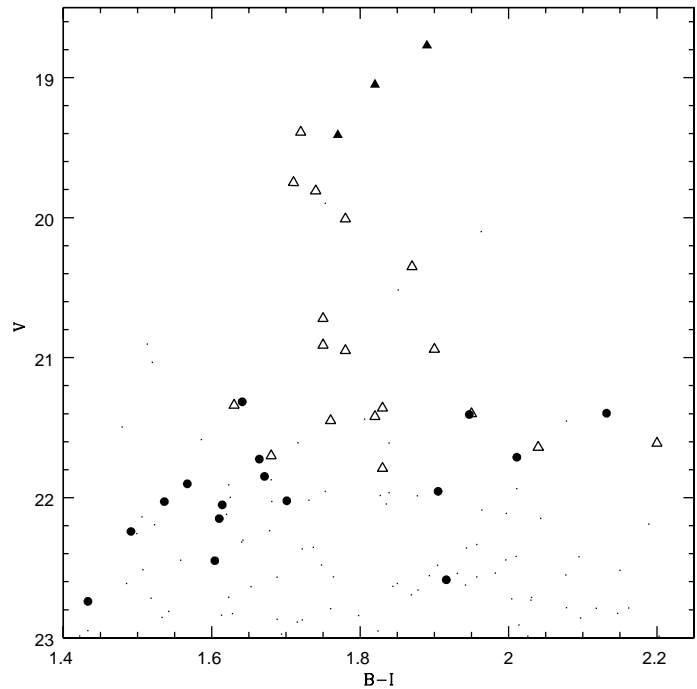}
\caption{Colour-magnitude diagram for the NGC 1052 and NGC 1316
GC systems. The NGC 1316 GCs are shown by triangles (filled
symbols when spectral 
age estimates are available). The NGC 1052 GCs are shown by
circles (large symbols when spectral ages are avilable). The NGC
1052 GC system reveals few if any young, bright GCs associated with
the central starburst of $\sim$2 Gyrs, unlike NGC 1316.  
}
\label{fig:1}       
\end{figure}

Proto-globular clusters certainly appear to be forming in
gas-rich mergers today such as the Antennae, but the question
remains `what contribution do GCs formed in recent mergers make
to the overall GC system of today's elliptical galaxies ?' An interesting
comparison is the cases of NGC 1316 and NGC 1052. Both are
thought to be 1--3 Gyr old merger remnants. NGC 1316 has a very
disturbed morphology with extensive gas and dust. NGC 1052 has an
elliptical appearance with some fine structure (Schweizer \&
Seitzer 1992), and evidence for HI gas infalling onto the nucleus
and in extended tails (van Gorkom et al. 1986). The CMDs for the
GC systems on these two galaxies, placed at the same distance,
are shown in Fig. 5. The NGC 1316 GC system reveals several GCs
at intermediate colour and bright magnitudes of 18.5 $<$ V $<$ 21.5. 
The brighter three
are confirmed by spectra to be $\sim$ 3 Gyrs old from Goudfrooij et
al. (2001). NGC 1052 has no equivalent counterpart GCs (Forbes et
al. 2001); here the CMD distribution of GCs suggests they are
predominately old. Subsequent Keck spectra of 16 NGC 1052 GCs
confirm they are old despite the galaxy having a young central
stellar population of $\sim$ 2 Gyrs (Pierce et al. 2005).

\begin{figure}
\centering
\includegraphics[height=10cm, angle=-90]{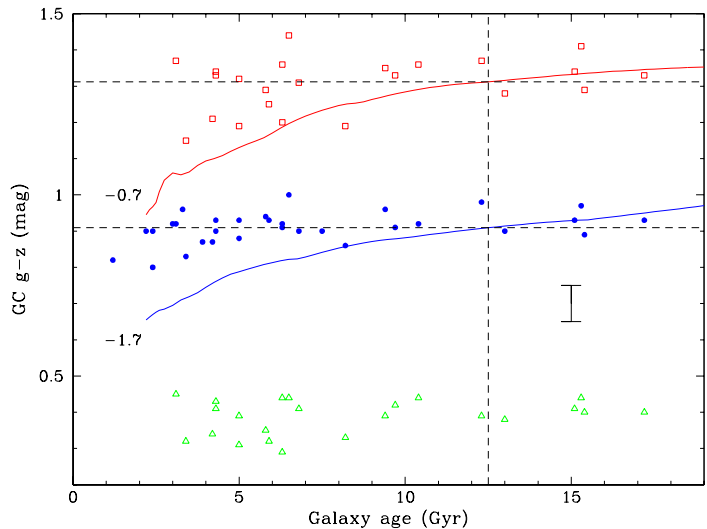}
\caption{Mean colour of GCs vs galaxy age for early-type 
Virgo galaxies. 
The dotted vertical line indicates an age of 12.5 Gyrs
(typical of Milky Way GCs). The solid lines show the change in
colour with age for two fixed metallicities, i.e. [Fe/H] = --0.7
and --1.7. 
The horizontal lines show the colour of a 12.5 Gyr population at
these two metallicities. The mean GC colours for the red, blue and
difference do not follow an evolutionary trend as expected if
they formed at the time of the galaxy starburst. Rather, they
have nearly constant colours consistent with them all having an
age of $\sim$12.5 Gyr. A typical error bar is shown.
}
\label{fig:1}       
\end{figure}

Many galaxies are found to have young central stellar
populations, indicative of an interaction or merger which induced
some star formation. If a significant number of GCs are formed
during this star formation episode we might expect the mean
colours of GCs to correlate with the time since the starburst. We
investigate this issue using early-type galaxies in the Virgo
cluster. Fig. 6 shows the mean colour of the red and blue GC
subpopulations, and their difference, with galaxy `age'. Colours
come from Peng et al. (2005) and galaxy ages from Caldwell et
al. (2003). The dotted vertical line indicates an age of 12.5 Gyrs
(typical of Milky Way GCs). The solid lines show the change in
colour with age for two fixed metallicities, i.e. [Fe/H] = --0.7
and --1.7 using models from Bruzual \& Charlot (2003). Where
these intersect the 12.5 Gyr line we draw a horizontal line to
guide the eye. The mean GC colours for the red/blue
subpopulations and
difference do not follow an evolutionary trend as expected if
they formed at the time of the galaxy starburst. Rather, they
have nearly constant colours consistent with them all having an
age of $\sim$12.5 Gyr. Thus any GCs formed in a recent merger 
associated
with the galaxy starburst appear to make little, if any, 
contribution to the overall GC system.

\section{Conclusions}

\noindent
$\bullet$ The ages, metallicities and alpha-element ratios of Milky Way GCs
measured with integrated spectra, and modified SSP models, are in
good agreement with CMD and higher resolution spectral results. \\
$\bullet$ Early formation (i.e. z $\ge$ 2, age older than 10
Gyrs) is favoured for most GCs.\\
$\bullet$ Recent accretion is responsible for some GCs in large
galaxies today.\\ 
$\bullet$ Recent mergers contribute few, if any, GCs to
today's elliptical galaxies. \\
$\bullet$ Reionization is the current best-bet for truncating the
formation of metal-poor GCs.\\
$\bullet$ The bimodality seen in GC specific frequency mirrors
the mass-to-light varations of the general galaxy population.\\

%
%
%
%

\printindex
\end{document}